\documentclass{elsartL}
\usepackage{graphicx}
\usepackage{amssymb}
\usepackage{amsmath}
\usepackage{physics}
\usepackage{mathrsfs}
\usepackage{mathtools}
\usepackage{rotating}
\usepackage{appendix}
\newcommand{\RNum}[1]{\uppercase\expandafter{\romannumeral #1\relax}}
\begin{document}

\begin{frontmatter}
\title{Modelling of the monthly global average temperature anomalies after the year 1900}
\author{Evangelos Matsinos}

\begin{abstract}
Two ways of parameterising the temporal dependence of the global average temperature anomalies after 1900 are put forth in this technical note. The models are fitted to the data of the Berkeley Earth Group 
(up to the end of 2023), after the application of corrections which translate the input so that its average value over the last five decades of the $19$-th century vanishes; as the corrections are evaluated 
separately for each month of the year, the seasonal effects are removed from the data. The results are given in a convenient form for use.\\
\noindent {\it PACS 2010: 88.05.Np; 92.30.Np; 89.60.-k; 92.60.Ry; 92.70.Aa; 92.70.Mn}
\end{abstract}
\end{frontmatter}

\section{\label{sec:Introduction}Introduction}

I am not aware of attempts at modelling the temporal dependence of the global average temperature anomalies (GATAs), other than those involving the application of locally-estimated (or -weighted) scatterplot 
smoothing filters to the data, i.e., the methods LOESS and LOWESS, respectively. I am equally unaware of attempts at applying suitable low-pass filters to the GATAs, a subject which I might address in a 
future study. The present paper relates to the mathematical modelling of the GATAs in terms of simple polynomials; two such models will be examined: the first relies on four connected line segments, whereas 
the second features a function with continuous first derivative with respect to time.

Bypassing the obvious academical interest in this subject, the temporal dependence of the annual GATAs, obtained from the twelve fitted values of the monthly GATAs in each year, is relevant to simulations 
of severe weather phenomena, hence to vital information about the appropriate strategy to cope with the effects of the climate change. In addition, it provides invaluable input to studies of such phenomena 
(e.g., relating to their frequency of occurrence and to their intensity) with the course of time, corresponding to the temporal interval during which the determination of monthly GATAs has been pursued. On 
the other hand, the \emph{central tendency} of the monthly GATAs, which is devoid of the short-term (i.e., characterised by a representative timespan which does not exceed the approximate period of the Solar 
activity) fluctuations of the original data, may be used in order to study (in a transparent way) systematic differences between months or sets of months: for instance, to determine whether the global 
absolute temperature increases (on average) faster or slower (at this time) for a given month (or sets of months) of the year (in comparison with other months or sets of months). Furthermore, the monthly 
GATAs are expected to be useful in simulations of weather phenomena which demonstrate an annual seasonal pattern (seasonality), i.e., different frequency of occurrence and/or different intensity during the 
course of the year.

Input to this work are the monthly GATAs of the Berkeley Earth Group \cite{BerkeleyEarth} from the start of 1900 to the end of 2023. Given that this paper elaborates on an observation made in Ref.~\cite{Matsinos2023}, 
it may be thought of as a continuation of that study. On pp.~10-11 of Ref.~\cite{Matsinos2023}, one reads: ``\dots the GATAs remained close to (or below) the baseline temperature levels until about 1910, 
then gradually increased (by about $0.3^\circ$C) until about the early 1940s. A rough plateau may be seen during the subsequent three decades, after which the GATAs have systematically increased (by about 
$0.2^\circ$C per decade) until the present time.''

The structure of this paper is as follows. Section \ref{sec:Method} presents the theoretical background which is relevant to the data analysis. Section \ref{sec:Results} provides the details of the analysis 
and the results emerging thereof. Finally, Section \ref{sec:Conclusions} summarises the findings of this work. To improve readability, the figures are placed at the end of the paper. A few Excel files, 
containing the important results obtained from the data, are made available as ancillary material.

In view of the facts that the input data to this study comprise global average temperatures and that the grid points, at which the absolute temperatures are measured or inferred, are evenly distributed 
between the two hemispheres, the notion of seasons is almost~\footnote{The dominant source of asymmetry is the different distribution of land between the two hemispheres.} devoid of meaning: for that reason, 
any reference to seasons will be avoided.

\section{\label{sec:Method}Methods}

\subsection{\label{sec:Modelling}On modelling the temporal dependence of the global average temperature anomaly after 1900}

\begin{figure}
\begin{center}
\includegraphics [width=15.5cm] {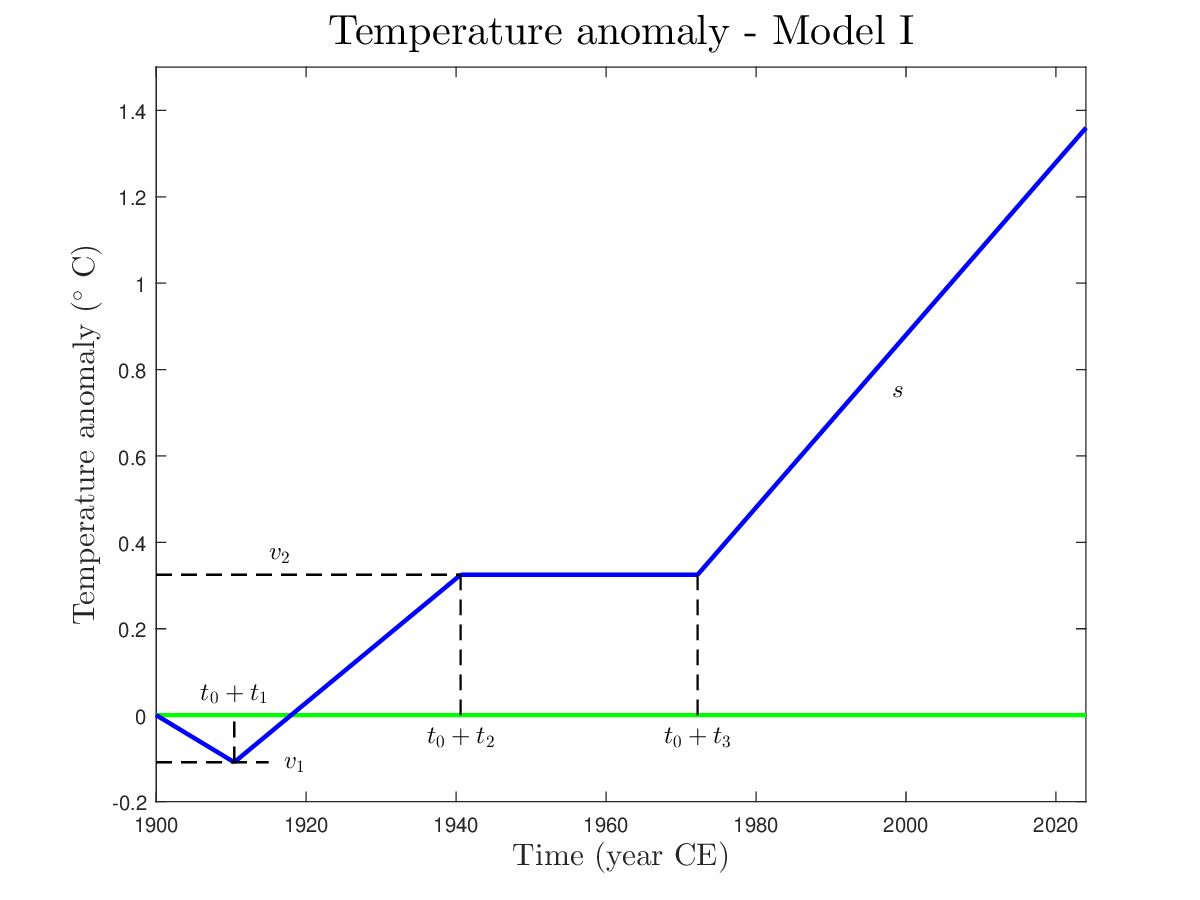}
\caption{\label{fig:Model1}The parameterisation of the temporal dependence of the monthly GATAs after $t_0 = 1900$ CE with Model \RNum{1} involves four connected line segments, see Eqs.~(\ref{eq:EQ001}). 
The derivative of this function is discontinuous at the origin ($t = t_0$) and at $t=t_0 + t_{1,2,3}$, see text. The green horizontal solid straight line marks the pre-1900 temperature standard.}
\vspace{0.5cm}
\end{center}
\end{figure}

Where the modelling is concerned, it will be assumed that the clock starts ticking on 1 January 1900 CE, 00:00 UTC. All time instants $t \geq t_0$ are redefined according to the replacement $t \to t - t_0$, 
where $t_0 \coloneqq 1900$. This operation relates to the independent variable in the parameterisation of the temporal dependence of the monthly GATAs and, of course, to those of the parameters in the two 
models which represent time instants. Two modelling options will be put forth (and into application) in this work.
\begin{itemize}
\item Model \RNum{1} is straightforward: it features four connected line segments between $t_0$ and the `current' time, to be identified herein with the end of 2023, thus corresponding to $t=124$ years (yr), 
see Fig.~\ref{fig:Model1}. Three nodes ($t_1$, $t_2$, and $t_3$) will be identified as follows: the time instant $t_1 \approx 10$ yr corresponds to the observed minimum of the monthly GATAs around 1910; the 
domain between $t_1$ and $t_2 \approx 40$ yr is characterised by a moderate increase in the GATAs until they reach the plateau spanning about three decades (i.e., between $t_2$ and $t_3 \approx 70$ yr), see 
Fig.~5 of Ref.~\cite{Matsinos2023} as well as Appendix \ref{App:AppA} of this study. A sharp increase in the GATAs with time is evident after $t_3$. Although a continuous function is constructed by means of 
these line segments, its time derivative is evidently discontinuous at the nodes $t=t_{1,2,3}$; moreover, given that the monthly GATAs will refer to the `pre-industrial' (1850-1899) \cite{IPCC2018} temperature 
standard (baseline solution), the first derivative with respect to time is also discontinuous at $t=0$ (unless the slope of the first line segment, i.e., between $t=0$ and $t_1$, is forced to $0$, which is 
not the case herein). Model \RNum{1} achieves the description of the data on the basis of six parameters.
\item Model \RNum{2} features a continuous function, whose time derivative is (unlike Model \RNum{1}) also continuous in the entire time domain. In essence, the first two line segments of Model \RNum{1} are 
replaced in Model \RNum{2} by a quartic function with vanishing derivative at the two endpoints $t=0$ and $t_2$. The next two line segments of Model \RNum{1} are used with one modification, which enables a 
smooth transition between these segments, and thus removes the discontinuity of the first derivative of Model \RNum{1} at the node $t_3$; to this end, a suitable quadratic function in the vicinity of $t_3$ 
will be introduced. Model \RNum{2} achieves the description of the data on the basis of five parameters.
\end{itemize}

\subsubsection{\label{sec:Model1}Model \RNum{1}}

The parameterisation of the temporal dependence of the monthly GATAs $v$ in Model \RNum{1} is put in mathematical form as follows.
\begin{equation} \label{eq:EQ001}
v(t) = \left\{
\begin{array}{rl}
v_1 t /t_1 & \text{, if $0 \leq t \leq t_1$}\\
v_1 + (v_2-v_1) (t-t_1) / (t_2-t_1) & \text{, if $t_1 < t \leq t_2$}\\
v_2 & \text{, if $t_2 < t \leq t_3$}\\
v_2 + s (t-t_3) & \text{, if $t > t_3$}
\end{array} \right.
\end{equation}
The meaning of the six model parameters ($t_1$, $t_2$, $t_3$, $v_1$, $v_2$, and $s$) either has already been addressed or is self-explanatory from Fig.~\ref{fig:Model1}. As it represents the current rate of 
increase in the global average temperature, the slope $s$ is the most important parameter (in both models).

\subsubsection{\label{sec:Model2}Model \RNum{2}}

The parameterisation of the temporal dependence of the monthly GATAs in Model \RNum{2} starts with the replacement of the first two line segments of Model \RNum{1} with a quartic function, which was chosen 
such that one requirement be fulfilled in the context of mathematical minimalism: given that the selected function must have three extrema, a fourth-degree polynomial constitutes the simplest way towards the 
intended parameterisation. To prevent tiny fitted values for the parameters of the quartic function, another independent variable $\tilde{t}$ was introduced: $\tilde{t} \coloneqq t / 100$; the temporal 
interval from $t=0$ to time $t_2$ is thus expressed in centuries (abbreviated herein to `cent') rather than in years. The suggested parameterisation reads as:
\begin{equation} \label{eq:EQ002}
f(\tilde{t}) = a \tilde{t}^4 + b \tilde{t}^{\, 3} + c \tilde{t}^{\, 2} + d \tilde{t} + e \, \, \, ,
\end{equation}
where $a \neq 0$. Two constraints fix the model parameters $d$ and $e$ to $0$: $f(0)=f^\prime(0)=0$, where the prime indicates differentiation with respect to $\tilde{t}$. As the temperature anomalies refer 
to the average temperature over the last five decades of the $19$-th century, the first condition is imposed by continuity. The second constraint may be understood as emerging from the requirement that the 
details about the pre-1900 data be exclusively contained in the set of monthly average temperatures (and associated uncertainties), with no use of any further information from that era.

The derivative of the resulting quartic function reads as
\begin{equation} \label{eq:EQ003}
f^\prime (\tilde{t}) = 4 a \tilde{t}^{\, 3} + 3 b \tilde{t}^{\, 2} + 2 c \tilde{t} \, \, \, ,
\end{equation}
which yields the intended extremum at $\tilde{t}=0$ and two additional extrema at the roots of the quadratic equation
\begin{equation} \label{eq:EQ004}
4 a \tilde{t}^{\, 2} + 3 b \tilde{t} + 2 c = 0 \, \, \, .
\end{equation}
The condition for two distinct real roots is that the discriminant of the quadratic form of Eq.~(\ref{eq:EQ004}), $\mathscr{D} = 9 b^2 - 32 a c$, exceed $0$. In such a case, the roots are given by 
\begin{equation} \label{eq:EQ005}
\tilde{t}_{1,2} = \frac{-3 b \pm \sqrt{\mathscr{D}}}{8 a} \, \, \, .
\end{equation}
The larger root $\tilde{t}_e \coloneqq \max \{ \tilde{t}_{1,2} \}$ is associated with the quantity $t_2$ of Model \RNum{1} ($\tilde{t}_e = t_2 / 100$), whereas the smaller root is similarly associated with 
$t_1$. Of course, the abscissae of the extrema emerge from the values of the parameters $a$, $b$, and $c$ of Model \RNum{2}, they are not parameters themselves. Furthermore, the parameter $v_2$ of Model 
\RNum{1} is also not a parameter in Model \RNum{2}: owing to the continuity condition, the value of the plateau $v_2$ is simply equal to $f(\tilde{t}_e)$.

In addition, Model \RNum{2} features a smooth transition between the last two line segments of Model \RNum{1}. A transition interval $\Delta t$ must be defined (see Fig.~\ref{fig:Intersection}), chosen 
herein to be equal to $1$ yr. Provided that this window remains small ($\lessapprox 5$ yr), its exact choice is barely of significance where the important results are concerned.

In summary, the parameterisation of the temporal dependence of the monthly GATAs $v$ in Model \RNum{2} reads as follows.
\begin{equation} \label{eq:EQ006}
v(t) = \left\{
\begin{array}{rl}
f(\tilde{t}) & \text{, if $0 \leq \tilde{t} \leq \tilde{t}_e$}\\
f(\tilde{t}_e) & \text{, if $100 \tilde{t}_e < t \leq t_3-\Delta t/2$}\\
f(\tilde{t}_e) + \frac{s}{2 \Delta t} \left( t - t_3 + \frac{\Delta t}{2} \right)^2 & \text{, if $t_3-\Delta t/2 < t \leq t_3+\Delta t/2$}\\
f(\tilde{t}_e) + s (t - t_3) & \text{, if $t > t_3+\Delta t/2$}
\end{array} \right.
\end{equation}
where the function $f$ was defined in Eq.~(\ref{eq:EQ002}), with (as aforementioned) $d=e=0$.

Given that the supply of `good' initial guesses (required for the initialisation of the function-minimisation application) is indispensable to avoiding the entrapment of the application in local minima, a 
few words are due.

According to the observations of Ref.~\cite{Matsinos2023} (pp.~10-11), the two roots of Eq.~(\ref{eq:EQ005}) sum up to about $0.5$ cent. This suggests~\footnote{The given approximate equality involves only 
the numerical values; the parameters $a$, $b$, and $c$ bear different units: $a$ is expressed in units of ($^\circ \, {\rm C \,\, cent}^{-4}$), $b$ in ($^\circ \, {\rm C \,\, cent}^{-3}$), and $c$ in 
($^\circ \, {\rm C \,\, cent}^{-2}$).\label{ftn:FTN1}} that $2 a \approx - 3 b$. On the other hand, their absolute difference is approximately equal to $0.3$ cent. This observation finally leads to the 
approximate equality $16 c \approx -1.92 b$ (keep footnote \ref{ftn:FTN1} in mind). Owing to the fact that a negative curvature for $f(\tilde{t})$ is observed at $\tilde{t} = 0$, the parameters $a$ and $c$ 
are negative, whereas $b$ is positive. Finally, a good initial guess for the parameter $b$ may be obtained from the observation $f(\tilde{t}_e) \approx 0.3^\circ$C.

\subsection{\label{sec:MinimisationFunction}Minimisation function}

The large interannual fluctuation, which is readily noticeable in the sets of monthly GATAs, e.g., see Figs.~1-4, 6, and 7 of Ref.~\cite{Matsinos2023}, is partly due to (regular or chaotic) natural causes 
(e.g., the Solar cycle, `El Ni$\tilde{\rm n}$o'- and `La Ni$\tilde{\rm n}$a'-type of phenomena, the volcanic activity, etc.), and partly due to the characteristics of chaotic systems, as the weather seems 
to be \cite{Shen2021}. Furthermore, the sharp decrease in the reported uncertainties of the monthly GATAs in the 1950s and early 1960s exacerbates one's uneasy feeling of large, unexplained variation which 
the visual inspection of the monthly GATAs imparts~\footnote{As the statistical uncertainties of the monthly GATAs decreased during the past $174$ years of diligent record-keeping, the interannual 
fluctuation, which is present in the data, became more pronounced (and more significant) with time.}.

To obtain results which represent the central tendency of the monthly GATAs, the contributions of the outliers to the minimisation function must be suppressed or, at least, reduced to a level which renders 
them harmless to the optimisation. On the whole, this may be achieved via the application of hard or soft weights to the data in the optimisation: approaches which apply hard weights (to the standardised 
residuals $r_i$ of the datapoints) identify (by ways which are frequently debatable or even nebulous) and remove the outliers from the database; on the contrary, those which rest upon the application of 
soft (continuous) weights aim at reducing the contributions of the outliers, without removing any entries from the database. The principal advantage of the latter, data-driven approaches bears on the 
dynamical determination of the contributions of the input datapoints to the minimisation function, depending on the `instantaneous' distance of each entry to the corresponding fitted value at each iteration 
of the optimisation process.

My first attempt at describing data, which contained a sizeable fraction of outliers, involved the application of Robust Statistics and the employment of a logarithmic minimisation function \cite{Matsinos1997}. 
In the meantime, seven additional methods, detailed in Section 4.1 of Ref.~\cite{Matsinos2019}, see pp.~18-19, were implemented. Two of these methods, which were already established in the mid-1970s 
\cite{Andrews1974,Beaton1974}, seem to be suitable for application in this study: they both feature a \emph{constant} contribution from the distant (corresponding to $\lvert r_i \rvert \gtrapprox 4.2$ 
\cite{Andrews1974} or $4.7$ \cite{Beaton1974}) outliers to the minimisation function.

After testing the three aforementioned methods (i.e., the logarithmic minimisation function \cite{Matsinos1997} and the methods of Refs.~\cite{Andrews1974,Beaton1974}), I finally decided to base the analysis 
of the main part of the paper on the minimisation function of Ref.~\cite{Matsinos1997}, but also report the results of the application of the minimisation function with Tukey weights (henceforth, Tukey 
minimisation function) \cite{Beaton1974} (with the corresponding default value of the tuning parameter $k = 4.685$) in the Excel files which are uploaded as ancillary material; the results, obtained with 
the minimisation function of Ref.~\cite{Andrews1974}, are not much different to those extracted with the Tukey minimisation function. Overall, the method of Ref.~\cite{Matsinos1997} is more effective (than 
the two other methods) in yielding a symmetrical distribution of the standardised residuals (and no significant dependence of these residuals on time), i.e., a solution which is expected to represent closer 
the central tendency of the data in the problem under consideration. The logarithmic minimisation function may easily be implemented, in a way similar to the seven methods of Ref.~\cite{Matsinos2019}, 
by introducing the weight $W (z_i)$ for the $i$-th datapoint according to the formula
\begin{equation} \label{eq:EQ007}
W (z_i) = \frac{\ln (1 + z_i^2)}{z_i^2} \, \, \, ,
\end{equation}
where $z_i \equiv r_i$. In case of vanishing $z_i$, the weight is obtained in the limit $z_i \to 0$:
\begin{equation} \label{eq:EQ008}
W (0) \coloneqq \lim_{z_i \to 0} \frac{\ln (1 + z_i^2)}{z_i^2} = 1 \, \, \, .
\end{equation}

For the numerical minimisation, the MINUIT software package \cite{James} of the CERN library (FORTRAN version) was exclusively used (in FORTRAN-callable mode); detailed relevant information may be obtained 
from Ref.~\cite{Matsinos2023b}, see Appendices B and C therein. As a result of the correlations among the model parameters, in conjunction with the considerable variation which is present in the set of the 
analysed monthly GATAs, the method MINOS of the MINUIT software package frequently failed to yield results for the asymmetrical uncertainties of the parameters of the two models. Due to this setback, MINOS 
was removed from the set of the MINUIT commands. However, as the emphasis is placed herein on the modelling of the central tendency of the monthly GATAs, i.e., not on the extraction of reliable uncertainties 
for the model parameters, the lack of reliable fitted uncertainties is of little account. Furthermore, though the MIGRAD uncertainties will be reported (in the ancillary material), they will not contribute 
to any part of the analysis. Some attention was paid in this study to safeguarding against the possible entrapment of the MINUIT-based application in a local minimum (which may be attributed to the 
characteristics of the input, as well as of the modelling): one hundred fits \emph{per case} were performed, starting from randomly-selected points in the parameter space.

\subsection{\label{sec:Tests}Tests}

The null hypothesis (namely that the observations may be attributed to stochastic fluctuation) will be assumed untenable herein, whenever the associated p-value drops below $1$~\%; this threshold is accepted 
as the outset of statistical significance by most statisticians. The $5$~\% level of significance is routinely used in hypothesis testing in several branches of Science (outside Physics); most statisticians 
associate that threshold with the outset of \emph{probable} statistical significance.

\subsubsection{\label{sec:SymmetryTest}Testing the standardised residuals for symmetry}

The symmetry of the distributions of the standardised residuals about the expectation value of $0$ (i.e., \emph{not} about each corresponding median) will be tested by means of the Wilcoxon signed-rank test 
\cite{Wilcoxon1945}. Owing to the largeness of the populations, the Edgeworth approximation \cite{Kolassa1995} is reliable (and will be used).

\subsubsection{\label{sec:NormalityTest}Testing the standardised residuals for normality}

Numerous algorithms are available for testing the normality of distributions, e.g., see Refs.~\cite{Shapiro2015,DAgostino1990} and the works cited therein. The normality of the distributions of the 
standardised residuals will be tested herein by means of the Shapiro-Wilk normality test, which emerges from many power studies as ``the most powerful normality test'' \cite{MohdRazali2011}. Introduced in 
1965 \cite{Shapiro1965} for small samples (in the first version of the algorithm, a maximum of $50$ observations could be tested), the test was substantially extended (for application to large samples, 
certainly up to $5\,000$ observations, perhaps to even larger samples) in a series of studies by Royston \cite{Royston1982-1995}.

\section{\label{sec:Results}Results}

The input comprises the monthly GATAs of the Berkeley Earth Group \cite{Rohde2020}, available from \cite{BerkeleyEarth}. Included in the file `Monthly Global Average Temperature' under `Global Monthly 
Averages (1850 - Recent)' are also uncertainties ($95$~\% level of confidence) associated with their estimates~\footnote{Of course, the models of Section \ref{sec:Modelling} can also be fitted to the second 
set of monthly GATAs which contain measurement uncertainties (determined per datapoint), i.e., to the HadCRUT5 dataset, which is available from \cite{HadCRUT5} and was analysed in Ref.~\cite{Matsinos2023}. 
This subject is not pursued in this work.}. The file contains two sets of results, representing different ways of treating the locations of the Earth's surface which are covered by sea ice: the first set is 
obtained via an extrapolation from land-surface air temperature data (method A), whereas the second involves an extrapolation from sea-surface water temperature data (method B). The Berkeley Earth Group 
recommend the use of the results with method A \cite{BerkeleyEarth}: ``We believe that the use of air temperatures above sea ice provides a more natural means of describing changes in Earth's surface 
temperature.''

For the purposes of this paper, the monthly GATAs of the Berkeley Earth Group~\footnote{I fail to understand why the past monthly GATAs of the Berkeley Earth Group (as well as their uncertainties) keep on 
changing as new (i.e., current) data are added to their two sets. It is my thesis that the Berkeley Earth Group must explain how the inclusion of the September-December 2023 GATAs in their database could 
possibly affect, even by minute amounts, their baseline solution (1951-1980).}, downloaded from \cite{BerkeleyEarth} on 17 January 2024, were processed as described in Section 2.1 of Ref.~\cite{Matsinos2023}. 
Visual inspection of the data suffices to derive the conclusion that the warmest occurrences of \emph{all} twelve months appeared within the last eight years (2016-2023), and that seven (!) of these 
occurrences, all months between June and December, fell in 2023, the warmest year on record \cite{Copernicus}. (Incidentally, September 2023, the current record holder for the largest monthly GATA, was 
warmer than the `average September' of the baseline solution by about $1.88^\circ$C.)

When fitting the models of Section \ref{sec:Modelling} to the monthly GATAs, December and February emerge as the worst-described months for both models and both choices of the minimisation function 
(logarithmic \cite{Matsinos1997} or Tukey's \cite{Beaton1974}), see Fig.~\ref{fig:WorstDescribedMonth} for the description of the February data achieved with the logarithmic minimisation function. On the 
opposite side, July emerges as the best-described month, see Fig.~\ref{fig:BestDescribedMonth}. Focussing on the data description achieved with the logarithmic minimisation function, one obtains the results 
of Table \ref{tab:FitQuantities} where the quality of the fits is concerned, and of Table \ref{tab:FitParameters} for the average values and uncertainties of the parameters of Model \RNum{1}; in case of 
Model \RNum{2}, estimates for four of the parameters of Model \RNum{1} were obtained from the fitted values of the three parameters of the quartic function. Owing to the large interannual fluctuation, 
\emph{all} fits are \emph{unacceptable} at the $1$~\% level of significance; this remark must be borne in mind whenever addressing the quality of the data description in this work.

\vspace{0.5cm}
\begin{table}[h!]
{\bf \caption{\label{tab:FitQuantities}}}Results for some characteristics of the fits of the two models of Section \ref{sec:Modelling} to the monthly GATAs of the Berkeley Earth Group. The quantities 
$F_{\rm min}$, N$_{\rm eff}$, BF, and p denote the minimal values of the logarithmic minimisation function \cite{Matsinos1997}, the effective number of degrees of freedom (to be identified with the sum of 
the weights of Eq.~(\ref{eq:EQ007}) over the data, reduced by the number of fit parameters), the Birge factor (i.e., the square root of the ratio $F_{\rm min}$/N$_{\rm eff}$, see also Appendix C of 
Ref.~\cite{Matsinos2023b}), and the corresponding p-value of each fit. To account for the real-valued N$_{\rm eff}$ in this study, the reported p-values were obtained via a cubic interpolation in $\ln({\rm p})$, 
involving the quoted $F_{\rm min}$ and four successive integer values around N$_{\rm eff}$, i.e., from $\lfloor {\rm N_{eff}} \rfloor - 1$ to $\lfloor {\rm N_{eff}} \rfloor + 2$.
\vspace{0.25cm}
\begin{center}
\begin{tabular}{|l||c|c|c|c||c|c|c|c||}
\hline
Month & $F_{\rm min}$ & N$_{\rm eff}$ & BF & p & $F_{\rm min}$ & N$_{\rm eff}$ & BF & p\\
\hline
\hline
 & \multicolumn{4}{|c||}{Model \RNum{1}} & \multicolumn{4}{|c||}{Model \RNum{2}}\\
\hline
January & $210.56$ & $53.85$ & $1.98$ & $2.14 \cdot 10^{-20}$ & $210.89$ & $54.93$ & $1.96$ & $3.97 \cdot 10^{-20}$\\
February & $243.89$ & $45.90$ & $2.31$ & $8.61 \cdot 10^{-29}$ & $246.65$ & $45.65$ & $2.32$ & $2.23 \cdot 10^{-29}$\\
March & $215.21$ & $53.28$ & $2.01$ & $2.43 \cdot 10^{-21}$ & $218.47$ & $53.22$ & $2.03$ & $6.66 \cdot 10^{-22}$\\
April & $204.36$ & $54.54$ & $1.94$ & $3.53 \cdot 10^{-19}$ & $207.15$ & $54.85$ & $1.94$ & $1.53 \cdot 10^{-19}$\\
May & $203.10$ & $55.19$ & $1.92$ & $8.71 \cdot 10^{-19}$ & $204.09$ & $56.06$ & $1.91$ & $1.08 \cdot 10^{-18}$\\
June & $173.67$ & $61.91$ & $1.67$ & $1.53 \cdot 10^{-12}$ & $176.66$ & $61.93$ & $1.69$ & $5.73 \cdot 10^{-13}$\\
July & $156.44$ & $67.85$ & $1.52$ & $5.94 \cdot 10^{-9}$ & $165.46$ & $65.66$ & $1.59$ & $1.42 \cdot 10^{-10}$\\
August & $169.97$ & $63.15$ & $1.64$ & $9.82 \cdot 10^{-12}$ & $172.72$ & $63.37$ & $1.65$ & $4.50 \cdot 10^{-12}$\\
September & $173.79$ & $61.03$ & $1.69$ & $9.17 \cdot 10^{-13}$ & $175.36$ & $61.87$ & $1.68$ & $8.55 \cdot 10^{-13}$\\
October & $178.79$ & $60.74$ & $1.72$ & $1.46 \cdot 10^{-13}$ & $183.67$ & $60.18$ & $1.75$ & $2.01 \cdot 10^{-14}$\\
November & $199.85$ & $55.45$ & $1.90$ & $3.44 \cdot 10^{-18}$ & $200.79$ & $56.11$ & $1.89$ & $3.74 \cdot 10^{-18}$\\
December & $247.13$ & $46.55$ & $2.30$ & $3.94 \cdot 10^{-29}$ & $248.00$ & $47.94$ & $2.27$ & $8.92 \cdot 10^{-29}$\\
\hline
\end{tabular}
\end{center}
\vspace{0.5cm}
\end{table}

\vspace{0.5cm}
\begin{table}[h!]
{\bf \caption{\label{tab:FitParameters}}}Results for the expectation values of the parameters which are associated with model \RNum{1}. The central values, which are quoted in the second column and those of 
the quantities $t_3$ and $s$ in the third, represent average fitted values over the twelve months of the year. The uncertainties represent the corresponding standard deviation (i.e., the square root of the 
unbiased variance); the fitted (MIGRAD) uncertainties were not used. For the sake of comparison, estimates for the quantities $t_1$, $t_2$, $v_1$, and $v_2$, obtained from the fitted values of the parameters 
$a$, $b$, and $c$ of Model \RNum{2}, see Section \ref{sec:Model2}, are given in the third column. In those cases, the reported results represent average values of the so-obtained estimates over the twelve 
months of the year (whereas the uncertainties represent the standard deviation of these estimates). In summary, the results obtained with the two parameterisations of Section \ref{sec:Modelling} agree within 
the quoted uncertainties. In particular, the values of the quantities $t_3$, $v_2$, and $s$, which exclusively determine the temporal dependence of the GATAs during the past six decades, come out nearly 
identical for the two parameterisations of Section \ref{sec:Modelling} (which, of course, is hardly surprising), an outcome which adds credibility to the predictions of this work regarding the near-future 
evolution of the monthly and annual GATAs.
\vspace{0.25cm}
\begin{center}
\begin{tabular}{|l|c|c|}
\hline
Parameter & Model \RNum{1} & Model \RNum{2}\\
\hline
\hline
$t_1$ (yr) & $8.4(3.0)$ & $12.4(3.6)$\\
$t_2$ (yr) & $38.7(5.5)$ & $40.3(6.0)$\\
$t_3$ (yr) & $71.1(4.6)$ & $71.2(4.4)$\\
$v_1$ ($^\circ$C) & $-0.163(70)$ & $-0.065(58)$\\
$v_2$ ($^\circ$C) & $0.284(51)$ & $0.286(51)$\\
$s$ ($^\circ$C yr$^{-1}$) & $0.0198(18)$ & $0.0198(19)$\\
\hline
\end{tabular}
\end{center}
\vspace{0.5cm}
\end{table}

The temporal dependence of the fitted values of the monthly GATAs for the two parameterisations of Section \ref{sec:Modelling} is shown in Figs.~\ref{fig:TAModel1} and \ref{fig:TAModel2}; as only these 
twelve sets of fitted values will be used henceforth, the short-term fluctuations, which were present in the original input, are suppressed. Average values for each year (and corresponding uncertainties) 
were obtained from the fitted results for the two parameterisations of Section \ref{sec:Modelling}; these average values (which will be referred to as `annual GATAs' henceforth) are reliable predictors of 
the representative GATA within the given year. Plots of the annual GATAs~\footnote{In the language of the `Analysis of Variance' (ANOVA) method, the sum of the monthly variations would be identified with 
the \emph{within-treatments} variation, whereas the uncertainties of the annual GATAs would be associated with the \emph{between-treatments} variation.} are given in Figs.~\ref{fig:ATAModel1} and 
\ref{fig:ATAModel2} for Models \RNum{1} and \RNum{2}, respectively; there is no room for misinterpretation.

The results of the tests for symmetry and normality of the distributions of the standardised residuals can be found in Table \ref{tab:Tests}. In all cases, no indication of asymmetry in these distributions 
about the expectation value of $0$ can be substantiated. On the contrary, the tests for normality fail for the five months between June and October: the distributions of the standardised residuals generally 
have longer tails than the normal distribution; this is verifiable via visual inspection, as well as after consideration of the kurtosis of the distributions of the standardised residuals (see ancillary 
material). This observation alone suffices to render the application of the conventional linear regression questionable and potentially unreliable in this study, calling for the employment of the robust 
methods of Ref.~\cite{Matsinos2019} (in particular of the methods of Andrews \cite{Andrews1974} or of Tukey \cite{Beaton1974}) or (even better, as the case turned out to be) of the logarithmic minimisation 
function \cite{Matsinos1997}.

\vspace{0.5cm}
\begin{table}[h!]
{\bf \caption{\label{tab:Tests}}}The p-values of the tests for symmetry and normality of the distributions of the standardised residuals for the two parameterisations of Section \ref{sec:Modelling}. The 
tests for symmetry (about the expectation value of $0$) were performed by means of the Wilcoxon signed-rank test \cite{Wilcoxon1945}, whereas those for normality were carried out by means of the Shapiro-Wilk 
test \cite{Shapiro1965,Royston1982-1995}. In the former case, the Edgeworth approximation \cite{Kolassa1995} was used and the (inessential) continuity correction was applied.
\vspace{0.25cm}
\begin{center}
\begin{tabular}{|l||c|c||c|c||}
\hline
Month & Symmetry & Normality & Symmetry & Normality\\
\hline
\hline
 & \multicolumn{2}{|c||}{Model \RNum{1}} & \multicolumn{2}{|c||}{Model \RNum{2}}\\
\hline
January & $6.32 \cdot 10^{-1}$ & $4.03 \cdot 10^{-1}$ & $6.22 \cdot 10^{-1}$ & $4.15 \cdot 10^{-1}$\\
February & $9.83 \cdot 10^{-1}$ & $1.70 \cdot 10^{-1}$ & $8.36 \cdot 10^{-1}$ & $1.64 \cdot 10^{-1}$\\
March & $1.96 \cdot 10^{-1}$ & $2.53 \cdot 10^{-1}$ & $5.16 \cdot 10^{-1}$ & $2.33 \cdot 10^{-1}$\\
April & $1.82 \cdot 10^{-1}$ & $1.26 \cdot 10^{-1}$ & $3.78 \cdot 10^{-1}$ & $9.21 \cdot 10^{-2}$\\
May & $8.14 \cdot 10^{-1}$ & $1.65 \cdot 10^{-1}$ & $8.51 \cdot 10^{-1}$ & $1.48 \cdot 10^{-1}$\\
June & $2.12 \cdot 10^{-1}$ & $3.49 \cdot 10^{-6}$ & $5.42 \cdot 10^{-1}$ & $2.12 \cdot 10^{-6}$\\
July & $4.66 \cdot 10^{-1}$ & $7.76 \cdot 10^{-7}$ & $9.40 \cdot 10^{-1}$ & $2.96 \cdot 10^{-6}$\\
August & $8.73 \cdot 10^{-1}$ & $3.16 \cdot 10^{-4}$ & $8.98 \cdot 10^{-1}$ & $7.17 \cdot 10^{-4}$\\
September & $4.53 \cdot 10^{-1}$ & $6.21 \cdot 10^{-10}$ & $6.66 \cdot 10^{-1}$ & $1.20 \cdot 10^{-9}$\\
October & $7.68 \cdot 10^{-1}$ & $7.93 \cdot 10^{-4}$ & $7.91 \cdot 10^{-1}$ & $4.40 \cdot 10^{-3}$\\
November & $3.28 \cdot 10^{-1}$ & $2.40 \cdot 10^{-1}$ & $3.19 \cdot 10^{-1}$ & $2.73 \cdot 10^{-1}$\\
December & $6.84 \cdot 10^{-2}$ & $2.56 \cdot 10^{-2}$ & $5.71 \cdot 10^{-2}$ & $1.99 \cdot 10^{-2}$\\
\hline
\end{tabular}
\end{center}
\vspace{0.5cm}
\end{table}

One issue calls for clarification. As aforementioned, July emerges from the fits as the best-described month, though the p-values of the Shapiro-Wilk test indicate that the distribution of the standardised 
residuals for that month is far from normality, and the same distribution has a large (positive) excess kurtosis. On the contrary, December and February are the worst-described months, though the distributions 
of the standardised residuals for these two months are not incompatible with normality, and both have a small (positive) excess kurtosis. How can these observations be reconciled?

To advance an explanation, one must obtain the probability density function (PDF) of the standardised residuals in the aforementioned cases. The kernel density estimation (KDE) of a PDF is an established, 
non-parametric, smoothing method \cite{Rosenblatt1956,Parzen1962}, employing kernels as weights on the elements of the input array of the observations (namely of the standardised residuals in this case): 
although a Gaussian kernel is a common choice, the selection of the kernel type largely depends on the characteristics of the array of the observations under consideration, and is frequently 
arbitrary~\footnote{For the purposes of the KDE of a PDF, kernels of finite (e.g., uniform, triangular, Epanechnikov, quartic, triweight, tricube, cosine, etc.) or infinite (e.g., Gaussian, logistic, sigmoid, 
etc.) support may be used.}. A Gaussian kernel was used herein, along with the recommended bandwidth parameter (which constitutes the optimal choice for the Gaussian kernel and Gaussian underlying 
distribution), see Ref.~\cite{Silverman1986}, Eq.~(3.28) on p.~45.

I shall next focus on the results obtained with Model \RNum{2} and the logarithmic minimisation function; similar results were extracted in the three other cases. Referring to Fig.~\ref{fig:PDFs}, one 
notices that the estimated PDF for July (continuous red curve) is leptokurtic, e.g., compare that curve with the PDF corresponding to the Gaussian function $N(\mu,\sigma^2)$ (dashed red curve), with $\mu \approx -0.22$ 
and $\sigma^2 \approx 8.96$ (representing the sample mean value and sample variance for July); save for the right tail at $r_i \gtrapprox 6$, the PDF is undoubtedly not of Gaussian form. On the other hand, 
the estimated PDF for February (continuous blue curve) might be considered an approximate Gaussian function $N(\mu,\sigma^2)$ (dashed blue curve), with $\mu \approx 0.22$ and $\sigma^2 \approx 14.85$ (obtained 
from the sample of the standardised residuals for February). The two vertical dash-dotted straight lines delimit the domain, wherein the contribution from each datapoint to the minimisation function does 
not exceed $1$. The fraction of the datapoints within this domain is considerably larger for July; consequently, the contributions of the datapoints in the tails of the two PDFs to the minimisation function 
are sizeably larger for February. On the basis of these considerations, the substantially larger $F_{\rm min}$ values for February (compared with July) can be understood. In summary, there can be no doubt 
that more pronounced unexplained variation is present in the February GATAs, though (unlike July) that variation appears to be largely Gaussian in nature.

\section{\label{sec:Conclusions}Conclusions}

This technical note touched upon the parameterisation of the temporal dependence of the global average temperature anomalies (GATAs) after the year 1900. Two modelling options were put forth, see Section 
\ref{sec:Modelling}: the first relies on four connected line segments, whereas the second features a function with continuous first derivative in the entire time domain of this paper.

The two aforementioned models were fitted to the monthly GATAs of the Berkeley Earth Group from 1900 to the end of 2023, available from \cite{BerkeleyEarth}. To enable the extraction of solutions 
representing the central tendency of the monthly GATAs, two robust methods were employed: the first features a logarithmic minimisation function \cite{Matsinos1997}, whereas the second makes use of the Tukey 
minimisation function \cite{Beaton1974}. Although the results of this paper originate from the first method, those obtained with the second are also made available as ancillary material.

Attention was paid to the determination of the baseline solution, which is representative of the temperature standard (separately evaluated for each month of the year) in the pre-industrial era. As in the 
previous study \cite{Matsinos2023}, the recommendation by the Intergovernmental Panel on Climate Change (IPCC) was followed \cite{IPCC2018}: the baseline solution was obtained by averaging the monthly 
global average temperature in the second half of the $19$-th century, i.e., from the start of 1850 to the end of 1899.

The seasonality was eliminated, by removing from the input GATAs the month-dependent offset of the baseline solution. The uncertainties of the monthly GATAs (reported by the Berkeley Earth Group) after 
1900 were summed in quadrature with the corresponding uncertainties of the baseline solution, and the combined uncertainty was assigned to each datapoint. (The linear summation of the uncertainties, the 
`pessimistic approach' when combining uncertainties, might be pursued in a follow-up work.)

Fits to the resulting monthly GATAs were carried out, separately for each month, using the parameterisations of Section \ref{sec:Modelling} and the two choices of the minimisation function of Section 
\ref{sec:MinimisationFunction}. Owing to the largeness of their unexplained variation, the description of the GATAs, corresponding to the four-month interval from December to March, is inferior to that 
corresponding to the four-month interval from June to September. The best-described month is July, the month with the smallest unexplained variation.

The findings of this work complement the observation of Ref.~\cite{Matsinos2023} that the monthly GATAs decreased slightly between the years 1900 and about 1910, then increased moderately until about 1940. 
For the following three decades, the monthly GATAs do not exhibit a significant temporal dependence, whereas a sharp increase in the global average temperature (by about $0.2^\circ$C per decade) is observed 
from about 1970 onwards, see Table \ref{tab:FitParameters}.

The standardised residuals of the fits were subjected to tests for symmetry and normality by means of the Wilcoxon signed-rank test (see Section \ref{sec:SymmetryTest}) and of the Shapiro-Wilk test (see 
Section \ref{sec:NormalityTest}), respectively. The logarithmic minimisation function was proven effective in yielding symmetrical distributions; the tests for symmetry with the two parameterisations of 
Section \ref{sec:Modelling} never failed, see Table \ref{tab:Tests}. Although the distributions of the standardised residuals, corresponding to the seven-month interval from November to May, may be regarded 
as not incompatible with normality, those corresponding to the remaining part of the year undeniably have a positive excess kurtosis, suggesting longer tails (than the normal) in these distributions. The 
broadening of the distributions is (to a large degree) due to the sizeable decrease in the uncertainties of the monthly GATAs in the course of the 1950s and the early 1960s. In any case, these observations 
constitute justification for the use of robust methods in the analysis of the monthly GATAs.

The average of the fitted values, corresponding to the twelve months of a given year, is a representative estimate for the GATA of that year (annual GATA). Figures \ref{fig:ATAModel1} and \ref{fig:ATAModel2} 
display the temporal dependence of such average values for the two parameterisations of Section \ref{sec:Modelling}, suggesting that the potentially critical $1.5^\circ$ threshold above the pre-1900 
temperature standard \cite{IPCC2018} will be reached (assuming no significant departure from the current tendency) in the course of the next decade, see also Ref.~\cite{Matsinos2023}, in particular Appendix 
A therein. To be precise, the solutions, corresponding to the two parameterisations of Section \ref{sec:Modelling}, suggest that this crossing will occur in 2032; a similar result was obtained in 
Ref.~\cite{Matsinos2023} (from a slightly different analysis of the GATAs of the Berkeley Earth Group after 1980). Finally, the temporal dependence of the annual GATAs provides essential input to studies 
of the climate during the last $174$ years.

The central tendency of the monthly GATAs is devoid of the short-term fluctuations of the original data, and can be used in order to study systematic differences between months or `seasons', e.g., providing 
the means to determine whether the global absolute temperature increases faster or slower for a given month (or `season') of the year (in comparison with other months or `seasons'). The monthly GATAs may 
also be useful in simulations of weather phenomena which demonstrate an annual seasonal pattern.

Convenient Excel files, containing the complete set of results obtained in this study, have been uploaded as ancillary material.

\begin{ack}
The figures of this paper were created with MATLAB$\textsuperscript{\textregistered}$ (The MathWorks, Inc., Natick, Massachusetts, United States).

I have no affiliation with or involvement in any organisation, institution, company, or firm with/without financial interest in the subject matter of this paper, nor did I receive any compensation in any 
form to conduct this study.
\end{ack}

\clearpage
\newpage

\begin{figure}
\begin{center}
\includegraphics [width=15.5cm] {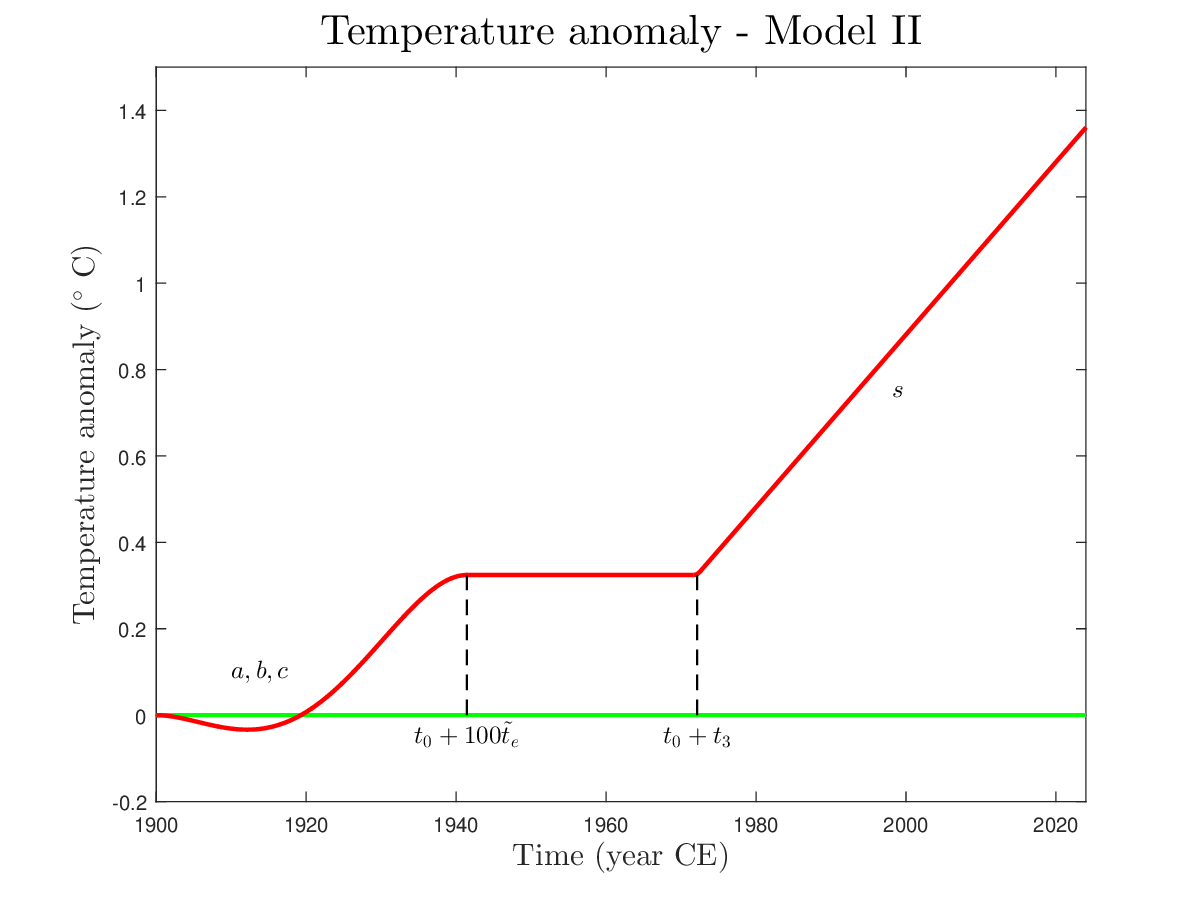}
\caption{\label{fig:Model2}The parameterisation of the temporal dependence of the monthly GATAs after $t_0 = 1900$ CE with Model \RNum{2} involves two suitable polynomials (one quartic, the other quadratic) 
and two line segments, see Eqs.~(\ref{eq:EQ006}). The derivative of this function is continuous in the entire time domain, see text and Fig.~\ref{fig:Intersection}. The green horizontal solid straight line 
marks the pre-1900 temperature standard.}
\vspace{0.5cm}
\end{center}
\end{figure}

\begin{figure}
\begin{center}
\includegraphics [width=15.5cm] {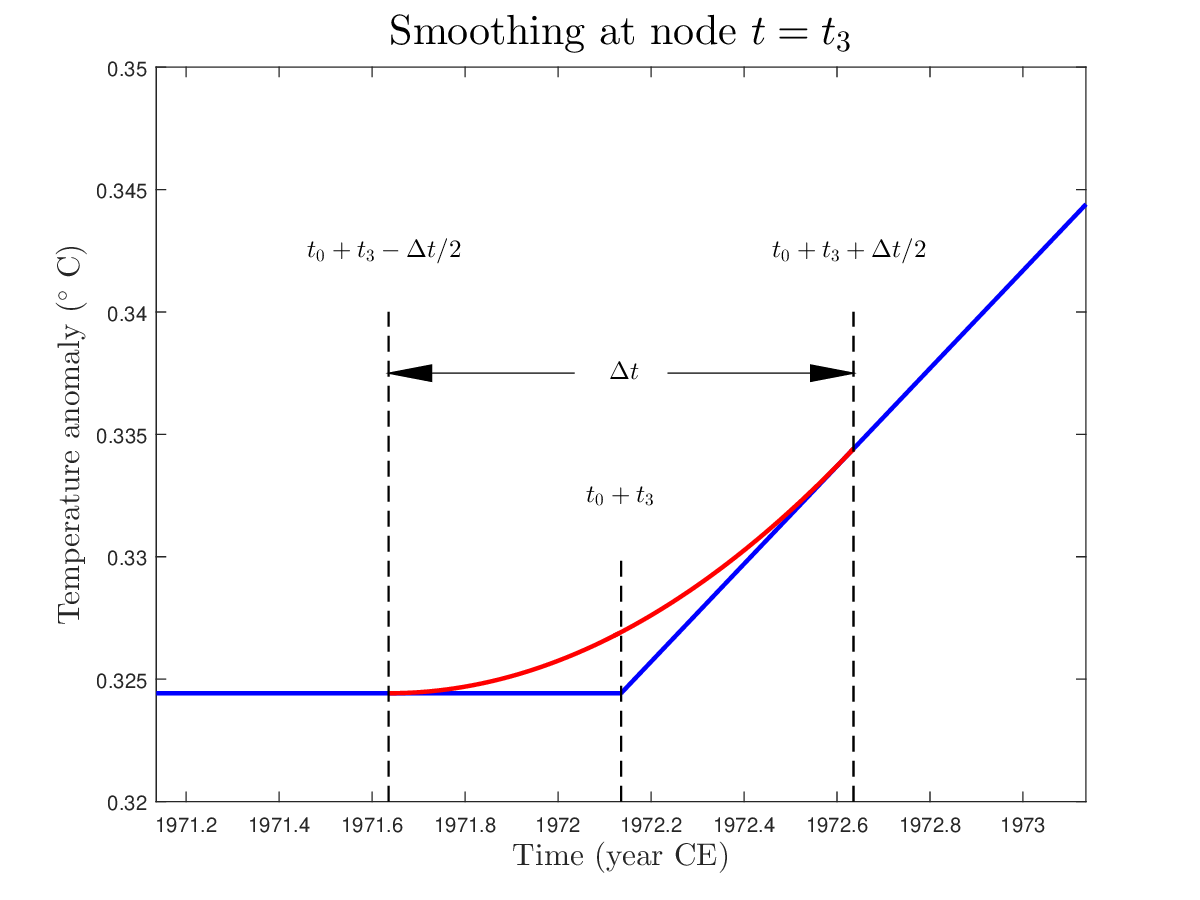}
\caption{\label{fig:Intersection}The smooth transition between the last two line segments of Model \RNum{1} is enabled in Model \RNum{2}. The blue solid straight lines of Model \RNum{1} are followed outside 
the domain which is delimited by the two vertical dashed straight lines. A quadratic form, selected in such a way that the derivative of $v(t)$ of Eqs.~(\ref{eq:EQ006}) be a continuous function, is followed 
within the aforementioned domain. In this work, the transition interval $\Delta t$ was chosen equal to $1$ yr, see comments in Section \ref{sec:Model2}.}
\vspace{0.5cm}
\end{center}
\end{figure}

\begin{figure}
\begin{center}
\includegraphics [width=15.5cm] {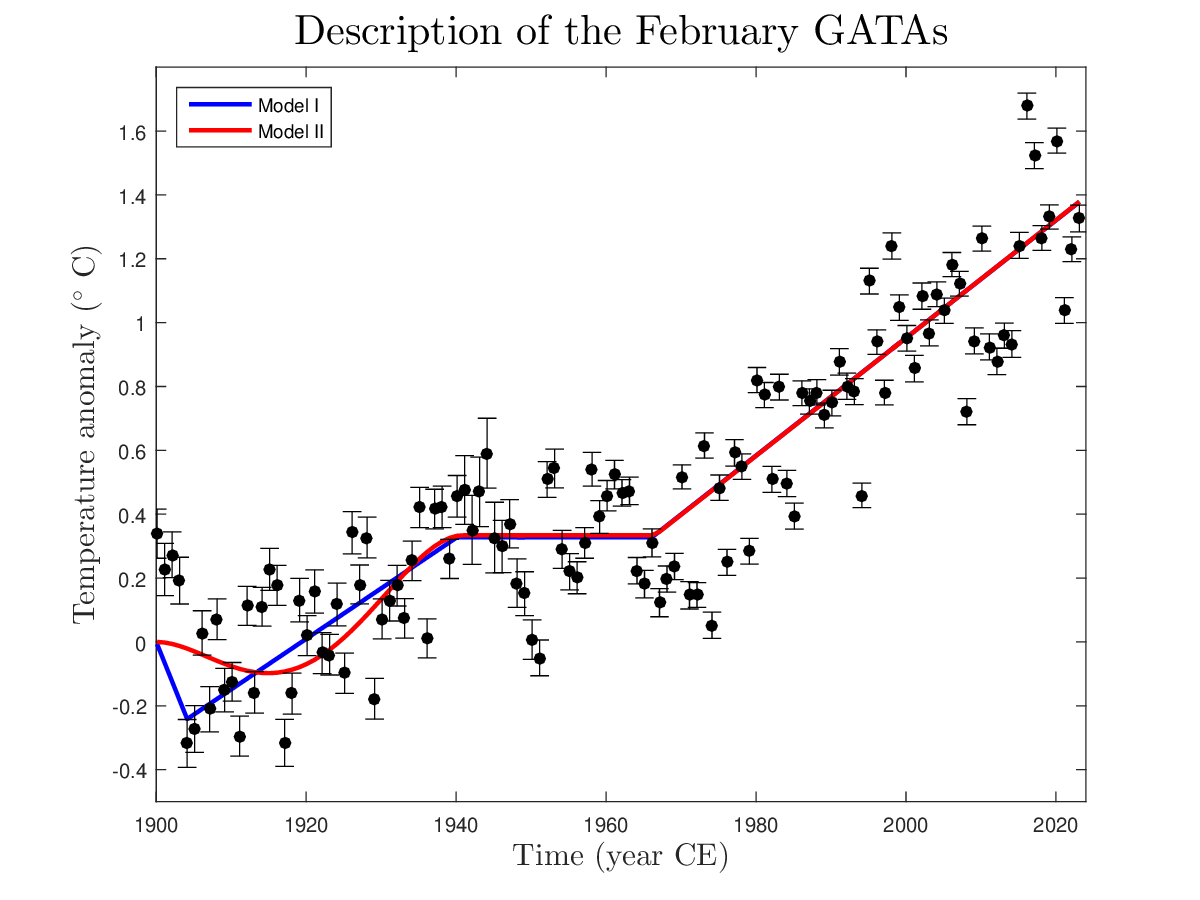}
\caption{\label{fig:WorstDescribedMonth}The description of the February GATAs: December and February emerge from the fits as the worst-described months. The fitted values, obtained with the two 
parameterisations of Section \ref{sec:Modelling} and the logarithmic minimisation function, are also shown.}
\vspace{0.5cm}
\end{center}
\end{figure}

\begin{figure}
\begin{center}
\includegraphics [width=15.5cm] {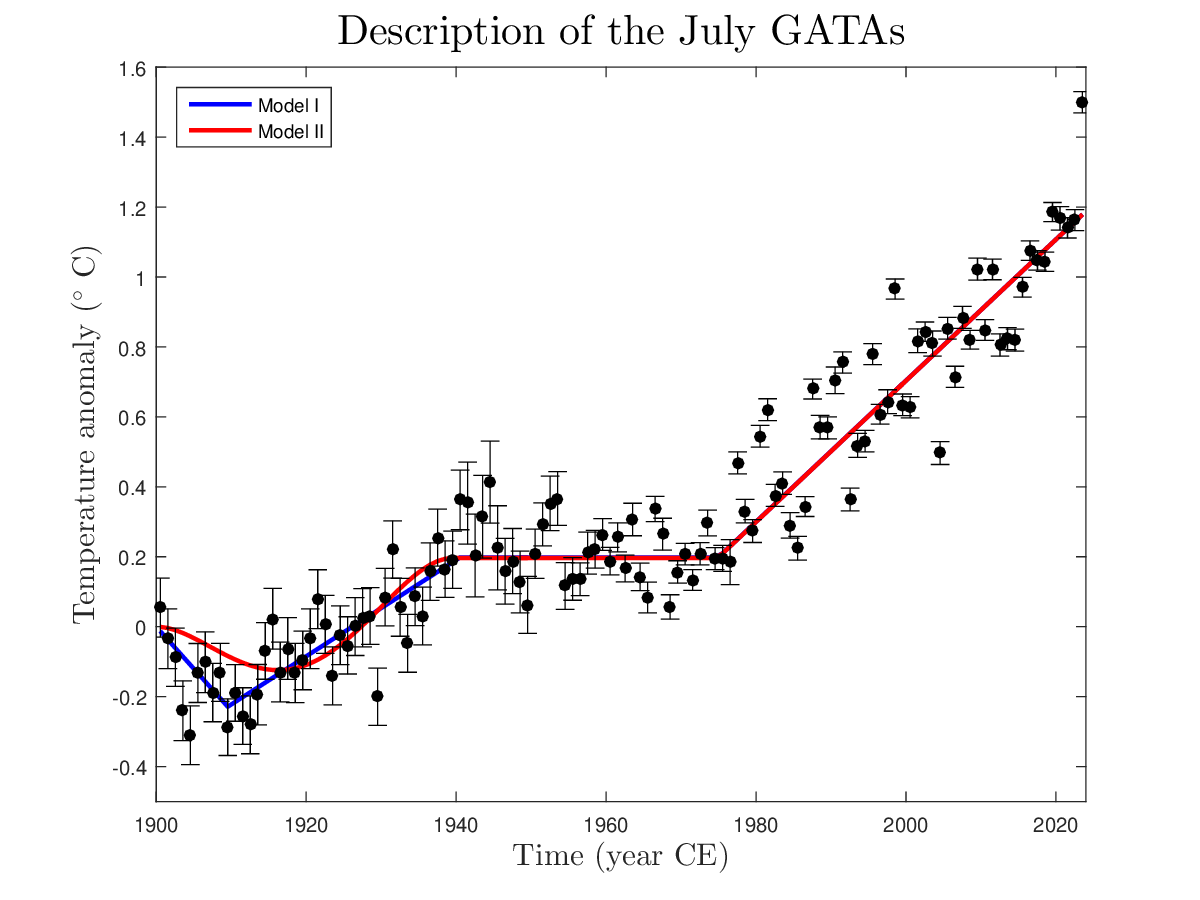}
\caption{\label{fig:BestDescribedMonth}The description of the July GATAs: July emerges from the fits as the best-described month. The fitted values, obtained with the two parameterisations of Section 
\ref{sec:Modelling} and the logarithmic minimisation function, are also shown.}
\vspace{0.5cm}
\end{center}
\end{figure}

\begin{figure}
\begin{center}
\includegraphics [width=15.5cm] {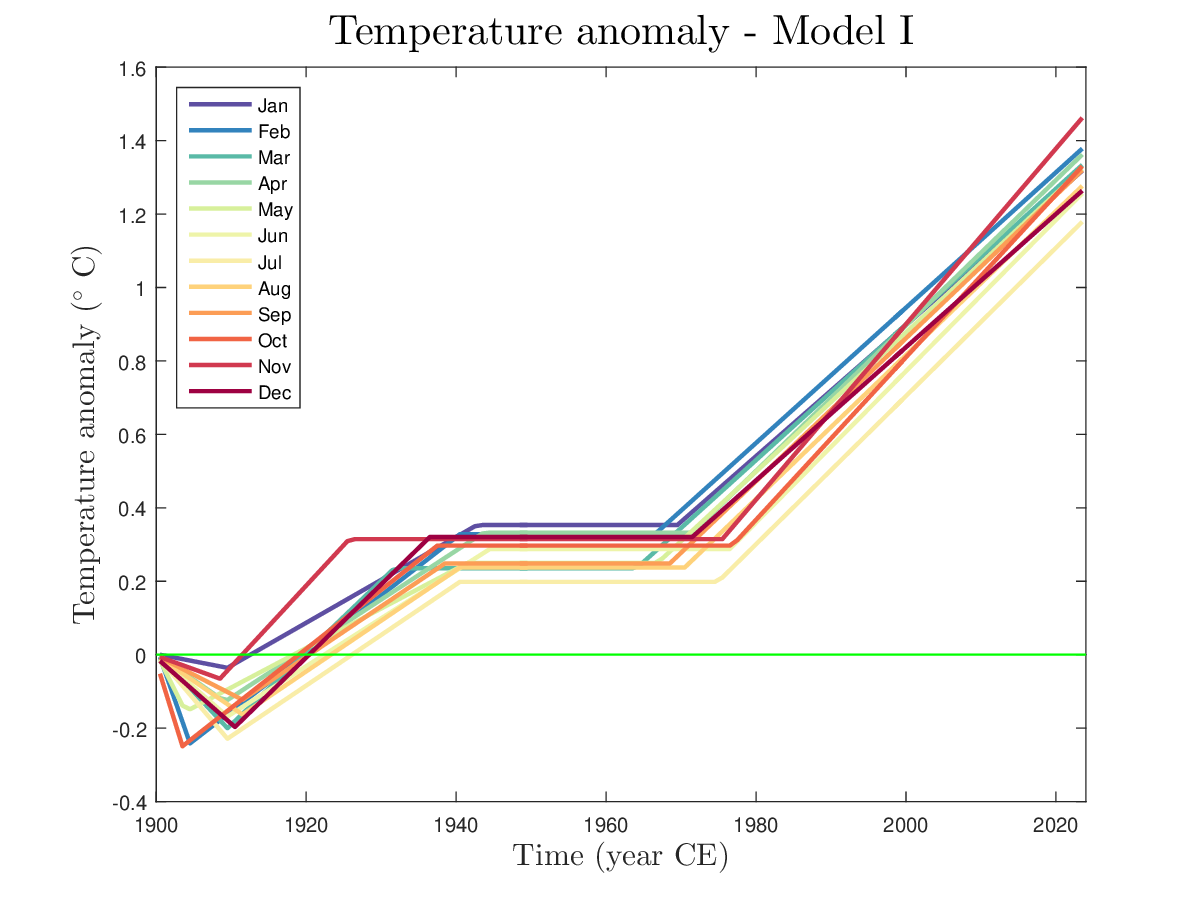}
\caption{\label{fig:TAModel1}The fitted values of the monthly GATAs of the Berkeley Earth Group \cite{BerkeleyEarth}, obtained with Model \RNum{1} (Section \ref{sec:Model1}) and the logarithmic minimisation 
function. The green horizontal straight line marks the pre-1900 temperature standard.}
\vspace{0.5cm}
\end{center}
\end{figure}

\begin{figure}
\begin{center}
\includegraphics [width=15.5cm] {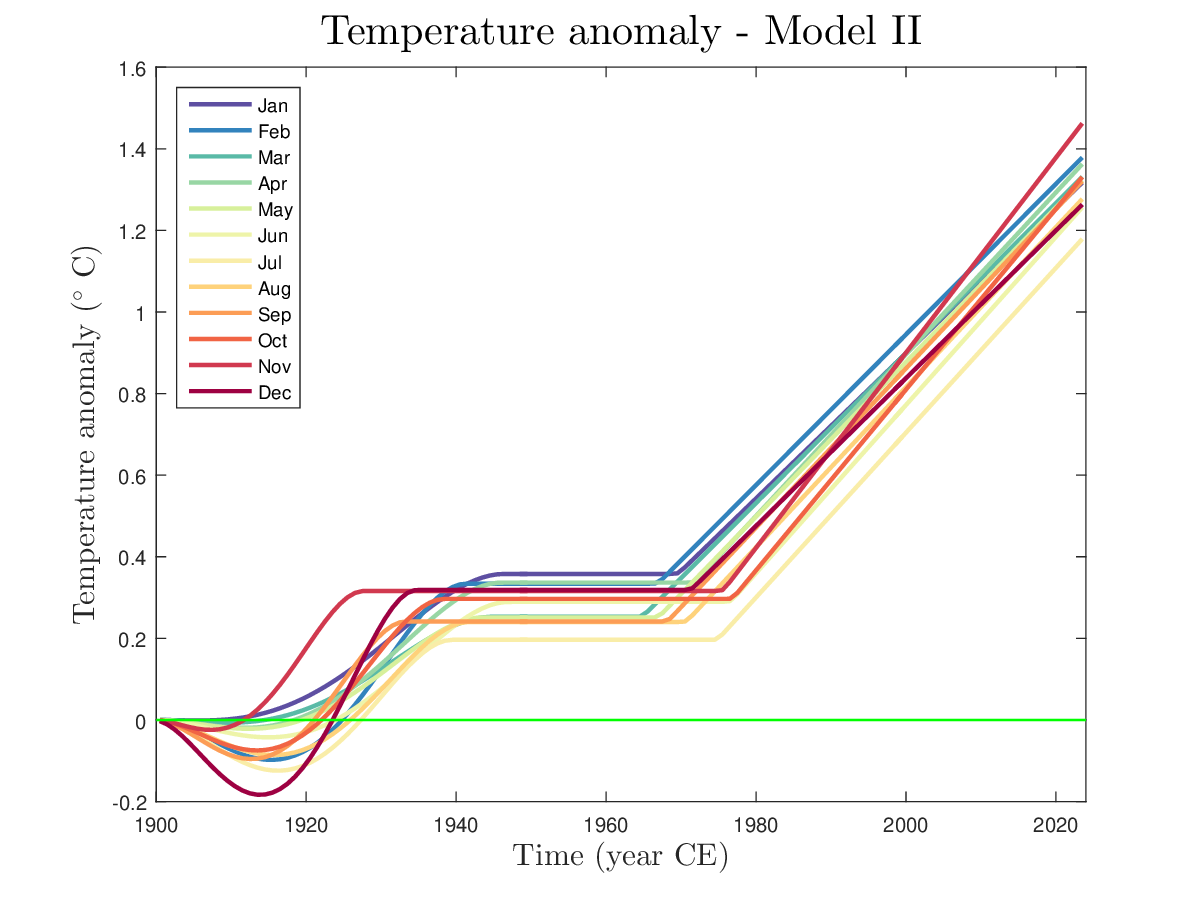}
\caption{\label{fig:TAModel2}The fitted values of the monthly GATAs of the Berkeley Earth Group \cite{BerkeleyEarth}, obtained with Model \RNum{2} (Section \ref{sec:Model2}) and the logarithmic minimisation 
function. The green horizontal straight line marks the pre-1900 temperature standard.}
\vspace{0.5cm}
\end{center}
\end{figure}

\begin{figure}
\begin{center}
\includegraphics [width=15.5cm] {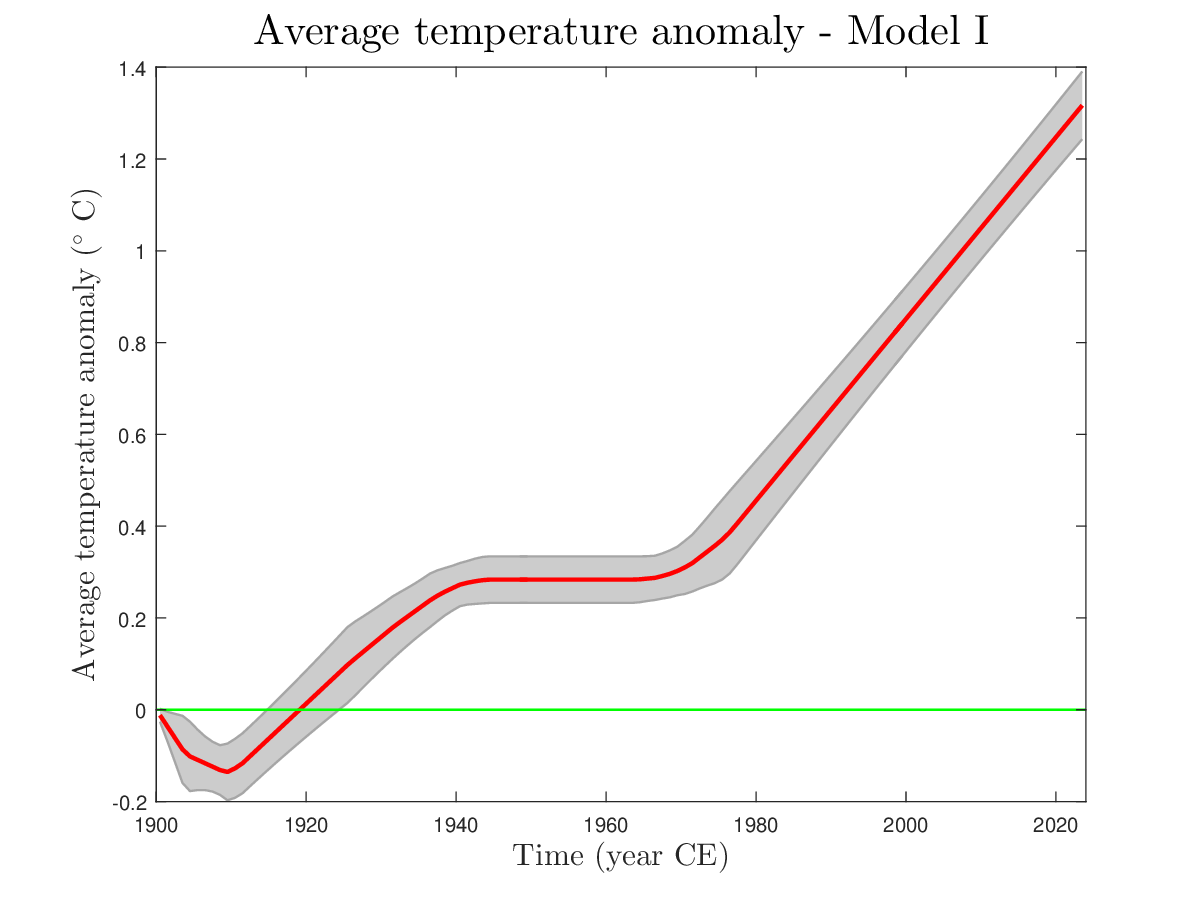}
\caption{\label{fig:ATAModel1}Annual GATAs (red curve), accompanied by $1 \sigma$ uncertainties (grey band), obtained with Model \RNum{1} (Section \ref{sec:Model1}) and the logarithmic minimisation function. 
The green horizontal solid straight line marks the pre-1900 temperature standard.}
\vspace{0.5cm}
\end{center}
\end{figure}

\begin{figure}
\begin{center}
\includegraphics [width=15.5cm] {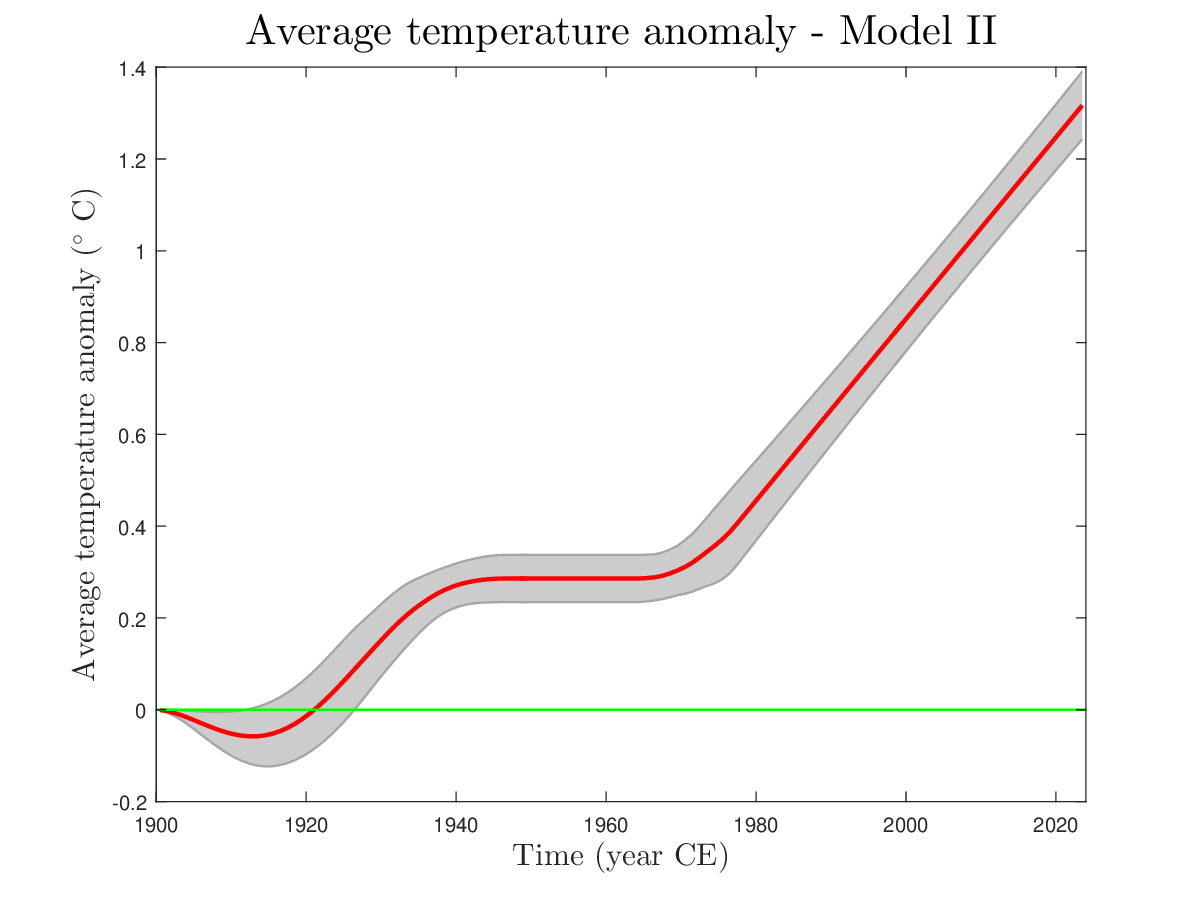}
\caption{\label{fig:ATAModel2}Annual GATAs (red curve), accompanied by $1 \sigma$ uncertainties (grey band), obtained with Model \RNum{2} (Section \ref{sec:Model2}) and the logarithmic minimisation function. 
The green horizontal solid straight line marks the pre-1900 temperature standard.}
\vspace{0.5cm}
\end{center}
\end{figure}

\begin{figure}
\begin{center}
\includegraphics [width=15.5cm] {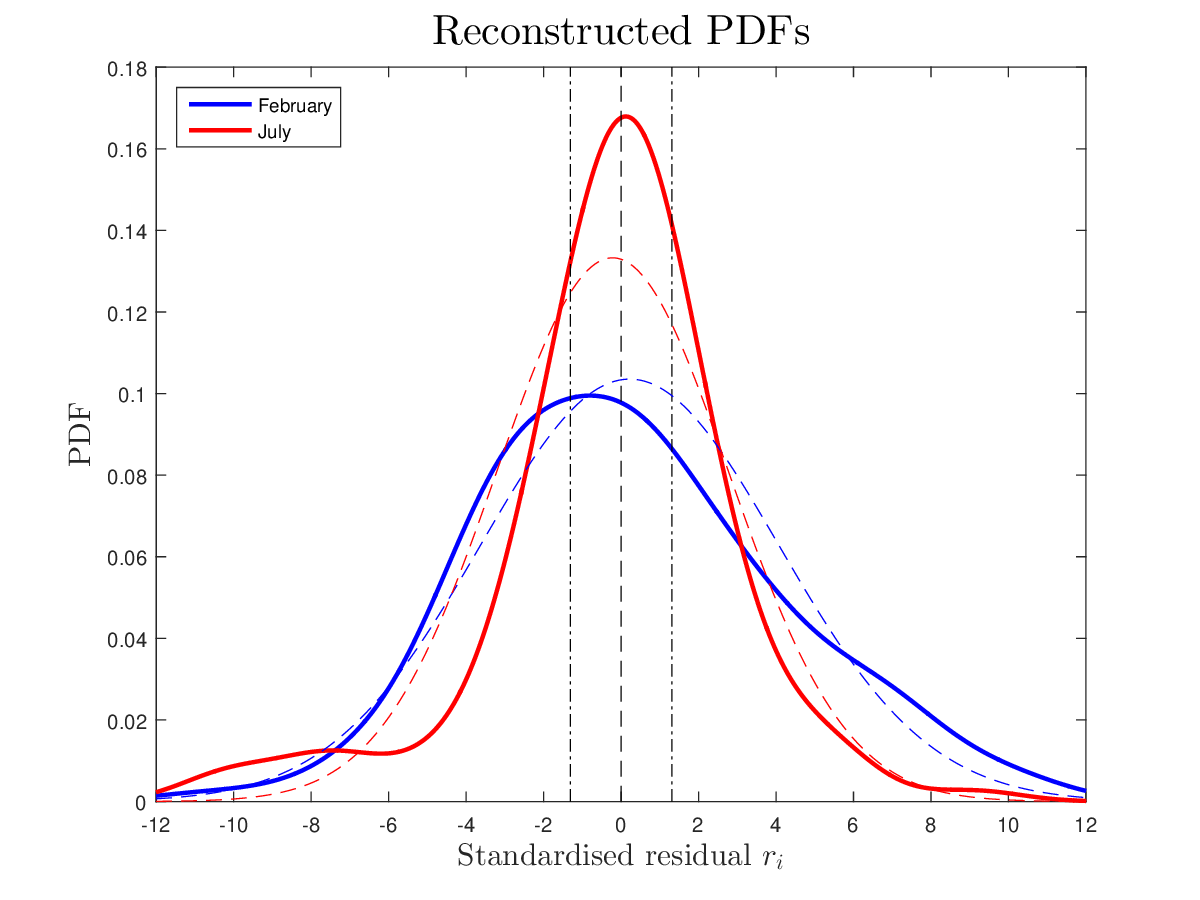}
\caption{\label{fig:PDFs}The PDFs of the standardised residuals for February (continuous blue curve) and July (continuous red curve); these functions were obtained via the method of kernel density estimation, 
featuring a Gaussian kernel, see text. For the sake of comparison, the two Gaussian functions $N(\mu,\sigma^2)$ are shown (dashed blue and red curves), with the mean values $\mu$ and the unbiased variances 
$\sigma^2$ obtained from the two sets of standardised residuals. A comparison of these PDFs demonstrates that (unlike July) the estimated PDF of the standardised residuals for February might be regarded as 
an approximate Gaussian function. The two vertical dash-dotted straight lines delimit the domain, wherein the contribution from each datapoint to the logarithmic minimisation function does not exceed $1$. 
The fraction of the datapoints within that domain is considerably larger for July, placing a smaller fraction of datapoints in the tails of the PDF for that month. In contrast, there can be no doubt that 
the February GATAs are subject to more pronounced unexplained variation, though (unlike July) that fluctuation appears to be largely Gaussian in nature.}
\vspace{0.5cm}
\end{center}
\end{figure}

\clearpage
\newpage
\appendix
\section{\label{App:AppA}The approximate constancy of the global average temperature anomalies during the middle part of the $20$-th century}

Had I been asked (before acquiring familiarity with the data relevant to this study), I would never have predicted the rough plateau of the GATAs between the early 1940s and the early 1970s. Given World War 
\RNum{2} and the post-war activity in rebuilding the destroyed infrastructure, I would have placed my bet on a palpable increase in the GATAs throughout the 1940s. Furthermore, there can be no doubt that 
most countries in North America and Europe experienced a remarkable economic growth during the 1950s and 1960s, based on a substantial rise in consumerism, propelled by the post-war euphoria in conjunction 
with the popularity of television~\footnote{The effectiveness of advertising on television, in persuading the general public that the product they see is the product they need, may not be underestimated.}; 
unbridled consumerism is inevitably associated with the climate change \cite{Panizzut2021}. Last but not least, those times saw the manufacturing \emph{en masse} of affordable long-distance modes of transport 
(powered by fossil fuels), enabling (among other advancements) the modernisation of agriculture and the optimisation of the process of food production and distribution\footnote{This development came with a 
high toll: transport-related CO$_2$ emissions account for 20-25~\% of the global CO$_2$ emissions today \cite{IEA2020}, e.g., compare Figs.~1.11 (p.~55) and 3.16 (p.~154).}; during the course of the 
aforementioned three decades, the number of motorised vehicles worldwide quadrupled \cite{Smil1994}. (That `blissful' era came to an abrupt end with the 1970s energy crisis.)

One plausible resolution of the paradox rests upon the possibility that the climate is a delayed system (e.g., see Ref.~\cite{Samset2020} and the references cited therein), in that there is time lag between 
causes and effects. Of course, it will not be the most welcoming news, if the climate indeed behaved as a delayed system with time lag of about thirty years \cite{Tebaldia2013}. In such a scenario, even if 
humankind could abolish its suicidal approach towards the environment and could adopt ecologically-sustainable practices \emph{today} (unlikely as this possibility might appear at this moment), the effects 
of such a `paradigm shift' would become noticeable in three decades, i.e., at the approximate time when the GATAs (on average) would be on the verge of crossing the $2^\circ$C threshold above the pre-1900 
temperature standard \cite{IPCC2018}, see Figs.~6 and 7 of Ref.~\cite{Matsinos2023}. Up to that moment, the climate would continue to respond in accordance with the ecological footprint of humankind during 
the past three decades.

\end{document}